\documentclass[aps,twocolumn,amsmath,amssymb,floatfix]{revtex4}

\usepackage{graphicx}
\usepackage{dcolumn} 
\usepackage{bm}      
\usepackage{color}

\begin{document}

\title{Helical rotating turbulence. Part II. Intermittency, scale invariance and structures
}
\author{P.D. Mininni$^{1,2}$ and A. Pouquet$^{2,3}$}
\affiliation{$^1$Departamento de F\'\i sica, Facultad de Ciencias Exactas y
         Naturales, Universidad de Buenos Aires and CONICET, Ciudad 
         Universitaria, 1428 Buenos Aires, Argentina. \\
             $^2$Computational and Information Systems Laboratory, NCAR, 
         P.O. Box 3000, Boulder, Colorado 80307-3000, U.S.A. \\
             $^3$Earth and Sun Systems Laboratory, NCAR, P.O. Box 3000, 
         Boulder, Colorado 80307-3000, U.S.A.}
\date{\today}

\begin{abstract}
We study the intermittency properties of the energy and helicity cascades in 
two $1536^3$ direct numerical simulations of helical rotating turbulence. 
Symmetric and anti-symmetric velocity increments are examined, as well 
as probability density functions of the velocity field and of the 
helicity density. It is found that the direct cascade of energy to small scales
is scale invariant and non-intermittent, whereas the direct cascade of 
helicity is highly intermittent. Furthermore, the study of structure 
functions of different orders allows us to identify a recovery of isotropy of 
strong events at very small scales in the flow.
Finally, we observe the juxtaposition in space of strong laminar and persistent helical columns next to time-varying vortex tangles, the former being associated with the self-similarity of energy and the latter with the intermittency of helicity.
\end{abstract}
\maketitle

\section{Introduction}
Turbulence is often referred to as the last unresolved main problem of classical physics. The diversity of applications 
of turbulent flows, from geophysics and astrophysics to engineering, and the observed complexity and lack of predictability of such flows both make 
the problem difficult to tackle, or even to define. Part of this 
complexity is due to the fact that turbulence comes in intermittent 
``gusts,'' strong events that are scarce, but still more frequent 
that what could be expected if normally distributed. These gusts give rise 
to the well-known break down of scale invariance in the flow \cite{Frisch}.

Intermittency is a highly spatially and temporally localized phenomenon. 
It is believed to be associated only with a forward cascade of an ideal 
invariant (a quantity conserved in the inviscid case), corresponding to 
the transfer of this quantity towards small scales with constant flux 
as a result of the non-linear coupling between modes; it is not directly 
related with the dimensionality of the 
problem. As an example, the energy cascade from larger to smaller scales 
in isotropic and homogeneous three dimensional turbulence is intermittent 
\cite{Sreenivasan97}. Magnetofluids, i.e., conducting fluids where the 
velocity field is coupled to the magnetic field, are intermittent in two 
\cite{Sorriso00} and in three dimensions \cite{Mininni07}, and in both 
cases a direct cascade of energy takes place. These examples are in 
contrast with the two dimensional neutral fluid case, for which the 
conservation of vorticity leads to an inverse energy cascade (a transfer 
of energy to the large scales with constant flux). While the direct 
cascade of enstrophy in this case is intermittent, there is evidence 
that the inverse cascade of energy is scale invariant and probability 
density functions of velocity fluctuations are close to Gaussian 
\cite{Bernard06}. For the case of rotating turbulence, studies of 
intermittency are not numerous 
since they require high Reyolds numbers together with low Rossby numbers; moreover, since in this case energy undergoes 
both a direct and an inverse cascade \cite{Cambon89,Waleffe93,Smith05}, 
it is unclear {\it a priori} whether intermittency is to be expected in 
the small scales or not, although, until the present work (see Sect. \ref{sec:intermittency}), there was no documented example of self-similar direct energy cascade to our knowledge. In experiments of freely decaying rotating 
turbulence \cite{Seiwert08}, it was found that rotation significantly 
decreases intermittency, although strictly scale-invariant 
(non-intermittent) statistics were not found. Such experimental 
results are also in agreement with recent analyses of data stemming from
direct numerical simulations (DNS) of non-helical rotating turbulence \cite{Muller07,Mininni09b}. Other laboratory experiments \cite{Baroud03} reported an even more pronounced reduction of the intermittency when rotation is present.

Although intermittency is believed to take place at small scales, 
strong events can affect the dynamics of the large scales, specially 
in systems close to criticality. As an example, it was shown in
\cite{Hoyng93,Charbonneau01,Mininni04} that local fluctuations 
of the kinetic helicity $H_V={\bf u} \cdot \nabla \times {\bf u}$ with ${\bf u}$ the velocity, can explain phase and amplitude variations of 
the 22-years solar cycle. Also, intermittency is known to affect the 
transport of momentum in atmospheric surface layers \cite{Kulkarni99}.

Considering the large number of degrees of freedom typical in 
turbulence (typical mesoscale flows in the atmosphere of the Earth can have, for 
instance, about $10^{18}$ degrees of freedom), many of the approaches to 
turbulence rely on an assumed scale invariance to model the flow 
statistics at small (often unresolved) scales. The search for 
self-similar quantities in three-dimensional turbulence is a long-standing 
problem, which would relate its study with critical phenomena and the 
out-of-equilibrium statistics of systems with a large number of modes, 
and which would allow the use of tools from quantum field theory, 
condensed matter, and statistical mechanics.

In a previous paper, hereafter referred to as Paper I \cite{paper1}, we presented 
results from two massive numerical simulations of helical rotating 
turbulence. Rotation and helicity are present in many atmospheric 
phenomena, as e.g., supercell storms and tornadoes 
\cite{Lilly88,Kerr96,Markowski98}. The simulations confirmed the scaling 
laws for the energy and helicity spectra predicted in \cite{Mininni09}. 
The development of anisotropies in the flow, as well as scaling laws 
in the directions parallel and perpendicular to the axis of rotation 
were also studied, considering both global measurements of anisotropy 
as well as spectral quantities. In this paper we study the intermittent properties of both the direct cascade of energy and
 the direct cascade of helicity in helical rotating turbulence, using the data from the simulations presented in Paper I. 
A decomposition into directions parallel and perpendicular to the axis of 
rotation is used to study anisotropy. Details of this decomposition, 
as well as the definitions of increments, structure functions, and 
scaling exponents used to quantify the intermittency are given in 
Sect. \ref{sec:defs}. The results of the decomposition performed on the data stemming from the high resolution DNS are presented in Sect. \ref{sec:functions}, 
where the possible recovery of isotropy at small scales is discussed. 
Averaging over different directions and over time, up to 
$4.3\times10^{11}$ data points were used to obtain statistical evidence 
that the direct cascade of energy is scale invariant, while the direct 
cascade of helicity is not, as discussed in Sect. \ref{sec:intermittency}. 
This result is obtained from the study of structure functions as well as 
probability density functions of velocity and helicity increments (Sect. 
\ref{sec:histo}). Finally, the development of structures at large and 
small scales in the flow is discussed in Sect. \ref{sec:structures} and 
the conclusions are presented in Sect. \ref{sec:conclusions}.

\section{Velocity and helicity increments\label{sec:defs}}
\subsection{Increments}
To study intermittency in the direct cascade of energy, we will use 
longitudinal increments of the velocity field ${\bf u}$,
\begin{equation}
\delta u ({\bf x}, \mbox{\boldmath $\ell$}) = \left[ {\bf u} 
    ({\bf x}+\mbox{\boldmath $\ell$}) - {\bf u}({\bf x}) \right] \cdot 
    \frac{|\mbox{\boldmath $\ell$}|}{\mbox{\boldmath $\ell$}},
\end{equation}
where the increment $\mbox{\boldmath $\ell$}$ can be in any direction. 
Structure functions of order $p$ are then defined as
\begin{equation}
S_p(\mbox{\boldmath $\ell$}) = \left< \delta u ^p 
    ({\bf x}, \mbox{\boldmath $\ell$}) \right> ,
\label{eq:structurev}
\end{equation}
where the brackets denote spatial average over all values of ${\bf x}$. 
The structure functions depend on the direction of the increment, and 
no assumption about isotropy or axisymmetry has yet been made.

The $S_p$ structure functions are of interest because for $p=2$ they 
are related to the two-point correlation function of the velocity, and 
thus to 
the energy spectrum through the convolution theorem. Therefore, 
a power law behavior can be expected for scales corresponding to the 
inertial range. Also, for the isotropic and homogeneous case, $S_3$ is 
related to the energy flux and scales linearly with the increment 
$\ell$ in the inertial range \cite{Kolmogorov41}. As a result, if the 
direct cascade is scale invariant, for an isotropic and homogeneous 
flow $S_p(\ell) \sim \ell^{p/3}$, and departures of the exponents from 
this relation are a signature of intermittency.

The study of intermittency in the direct cascade of helicity has been 
less explored. In the isotropic and homogeneous case, one can study it using for example
 structure functions based on the 
helicity flux, which is a third order field \cite{Chen03,Chen03b}. In real space, this flux 
can be written in two different ways, which follow from the r.h.s. of the 
K\'arm\'an-Howarth theorem for the helicity \cite{Chkhetiani96,Kurien03}:
\begin{equation}
\left< \left[{\bf u}({\bf x})\cdot \frac{|\mbox{\boldmath $\ell$}|}
       {\mbox{\boldmath $\ell$}} \right] \left[\left({\bf u}({\bf x}) 
       \times {\bf u}({\bf x}+\mbox{\boldmath $\ell$}) \right) 
       \cdot \frac{|\mbox{\boldmath $\ell$}|}{\mbox{\boldmath $\ell$}} 
       \right] \right> = \frac{1}{15} \delta \ell^2 ,
\end{equation}
or in terms of structure functions of the velocity and vorticity \cite{Gomez00},
\begin{eqnarray}
\left< \delta u ({\bf x}, \mbox{\boldmath $\ell$}) 
  \left[\delta {\bf u}({\bf x}, \mbox{\boldmath $\ell$}) \cdot 
  \delta \mbox{\boldmath $\omega$}({\bf x}, \mbox{\boldmath $\ell$})
  \right] \right> - \nonumber \\
- \frac{1}{2} \left< \delta \omega ({\bf x}, \mbox{\boldmath $\ell$}) 
  \left(\delta {\bf u}({\bf x}, \mbox{\boldmath $\ell$}) \right)^2 \right> 
= -\frac{4}{3} \tilde \epsilon \ell \ ,
\end{eqnarray}
where $\tilde \epsilon$ is the helicity injection rate. Both relations are 
equivalent, as it is easy to see from $\delta \omega \sim \delta u/\ell$ 
\cite{Gomez00}. However, here we want structure functions that can be 
associated to the spectral scaling, and must therefore be based on second 
order quantities. There are two candidates that follow from the 
expressions in the time derivative appearing in the K\'arm\'an-Howarth 
theorem for helicity \cite{Chkhetiani96,Kurien03}:
\begin{equation}
\left< \left[ u_i({\bf x}+\mbox{\boldmath $\ell$}) - u_i({\bf x}) \right] 
       \left[ u_j({\bf x}+\mbox{\boldmath $\ell$}) - u_j({\bf x}) \right]
\right> ,
\end{equation}
or \cite{Gomez00}
\begin{equation}
\left< \delta {\bf u}({\bf x}, \mbox{\boldmath $\ell$}) \cdot 
       \delta \mbox{\boldmath $\omega$}({\bf x}, \mbox{\boldmath $\ell$}) 
       \right> \ .
\end{equation}
Both quantities are Galilean invariant, as is required for the structure 
functions to be well behaved \cite{Tsinober92}. The first quantity is a 
second order tensor with indices $i$, $j$, while the second is a scalar. 
For simplicity, here we use the second expression, although both are 
related to the antisymmetric part of the two-point correlation tensor 
for the velocity. Then, structure functions for the helicity can be 
defined as
\begin{equation}
{\cal H}_p (\mbox{\boldmath $\ell$}) = \left< \left[
       \delta {\bf u}({\bf x}, \mbox{\boldmath $\ell$}) \cdot 
       \delta \mbox{\boldmath $\omega$}({\bf x}, \mbox{\boldmath $\ell$}) 
       \right]^p \right> .
\label{eq:structureh}
\end{equation}
With this definition, for isotropic and homogeneous turbulence the 
assumption of scale invariance leads to ${\cal H}_p(\ell) \sim \ell^{p/3}$.
Note that ${\cal H}_1$ is  second-order in the velocity whereas $S_1$ is first order.

\subsection{Parallel and perpendicular directions}
The development of anisotropies in a rotating flow has been studied in 
experiments \cite{Jacquin90,Morize05,Staplehurst08} and in numerical 
simulations (see e.g, \cite{Smith99}). Anisotropy was shown to develop 
in a range of Rossby numbers such that non-linear interactions are not 
completely damped with the scrambling effect of inertial waves 
\cite{Cambon97}. The presence of rotation breaks down the isotropy of the 
flow, introducing a preferred direction. Energy, as a result of resonant 
triad interactions, is transferred preferentially towards modes in 
spectral space perpendicular to the axis of rotation 
\cite{Cambon89,Waleffe93,Cambon97}, although linear effects may also be 
relevant in the formation of the structures \cite{Davidson06}.

\begin{figure}
\includegraphics[width=8.6cm]{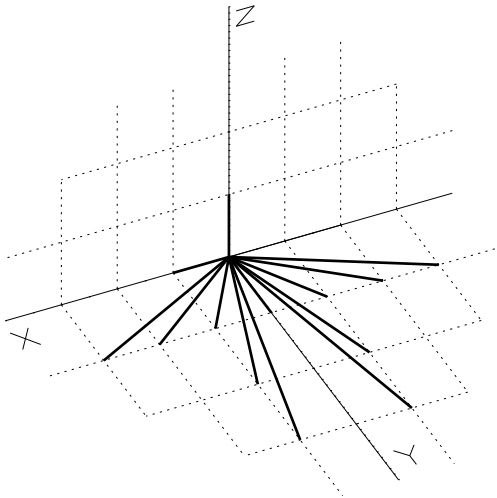}
\caption{The 12 generators used to compute increments in the $x$-$y$ 
plane, and the generator in the $z$ direction. The crossings of dotted 
lines indicate grid points in the numerical simulation.}
\label{fig:SO2} \end{figure}

In isotropic and homogeneous turbulence, it is a common practice to 
study velocity increments (or two-point correlation functions) in only a 
few directions, or to average over different directions as in, e.g., the 
leading (isotropic) term of the SO(3) decomposition 
\cite{Arad98,Biferale05}. Given the preferred direction in our problem, 
and the natural axisymmetry associated with it, we will be interested 
in increments parallel and perpendicular to the angular rotation 
$\mbox{\boldmath $\Omega$}$, which will be denoted respectively as 
$\ell_\parallel$ and $\mbox{\boldmath $\ell$}_\perp$. In principle, in the 
plane perpendicular to $\mbox{\boldmath $\Omega$}$, the increments 
$\mbox{\boldmath $\ell$}_\perp$ can be taken in every possible direction 
and later averaged to obtain structure functions that only depend on the 
scalar increment $\ell_\perp$. However, this requires interpolation of the 
three components of the velocity in the three dimensional space every 
time an increment does not 
reside on a computational grid point. To avoid 
the computational cost of this interpolation, we follow the procedure 
introduced in \cite{Taylor03} for the isotropic case. We only compute 
increments given by the product of an integer times a generator vector, 
with the generator vectors chosen such that they fill as uniformly as 
possible circles in the plane perpendicular to $\mbox{\boldmath $\Omega$}$, 
and such that their product with an integer always falls on a grid 
point.

Twelve generators were used in the $x$-$y$ plane: $(1,0,0)$, $(1,1,0)$, 
$(2,1,0)$, $(3,1,0)$, $(0,1,0)$, $(-1,1,0)$, $(1,2,0)$, $(-2,1,0)$, 
$(-1,2,0)$, $(1,3,0)$, $(-3,1,0)$, and $(-1,3,0)$ (in units of grid points 
in the simulation). These generators, plus the 12 generators obtained by 
multiplying them by $-1$ (or equivalently, considering negative increments) 
cover the plane in an approximately uniform way (see Fig. \ref{fig:SO2}). 
In the $z$ direction (the direction parallel to the axis of rotation) the 
generator for the increments is the vector $(0,0,1)$. Given these 
generators, increments in Eqs. (\ref{eq:structurev}) and 
(\ref{eq:structureh}) are created by multiplying the generators by integer 
numbers.  With these choices, all increments reside on grid points and no 
interpolation in the computation of the structure functions for each 
direction is required. Once structure functions for all directions have 
been computed, structure functions in the perpendicular direction 
$S_p(l_\perp)$ and ${\cal H}_p(l_\perp)$ are obtained by averaging over the 
results for the 12 directions in the $x$-$y$ plane. As the generators 
have different lengths, interpolation in this step is required, but it is 
less costly from the computational point of view as only interpolation 
of scalar one-dimensional functions is needed. The structure functions 
in the parallel direction, $S_p(l_\parallel)$ and ${\cal H}_p(l_\parallel)$, are 
obtained directly from the generator in the $z$ direction. Finally, 
average in time (using snapshots of the velocity field at different 
turnover times) can be computed.

In our case, we use for the run with 
the larger rotation rate ($\Omega=9$),
ten snapshots of the velocity field spanning ten turnover times, from 
$t=20$ to $30$ (see paper I). As a result, considering that each snapshot 
has $1536^3\approx 3.6\times 10^{9}$ grid points, and considering the 
twelve generators used, each increment in $S_p(l_\perp)$ and 
${\cal H}_p(l_\perp)$ results from an averaging over $4.3\times 10^{11}$ data points. 
In the case of $S_p(l_\parallel)$ and ${\cal H}_p(l_\parallel)$, each increment 
is obtained using $\approx 3.6\times 10^{10}$ data points.

\section{Structure functions\label{sec:functions}}
\subsection{Velocity structure functions}

\begin{figure}
\includegraphics[width=8.2cm]{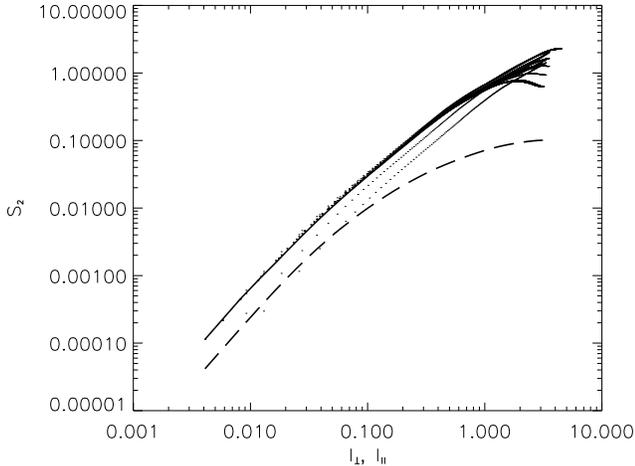}
\caption{Second order structure functions $S_2$ at $t=30$ in run B with $Ro=0.06$. The 
dotted lines indicate the different structure functions in the twelve 
directions given by the generators in the $x$-$y$ plane, and the thick 
solid curve is the average $S_2(\ell_\perp)$. The thick dashed curve 
corresponds to increments in the $z$ direction and is $S_2(\ell_\parallel)$.}
\label{fig:S2all} \end{figure}

Two simulations were used for the analysis, described in more detail in 
Paper I. One of the simulations (hereafter, run A), has $\Omega = 0.06$, 
a Rossby number $\textrm{Ro} \approx 8.5$, and was continued for almost 10 
turnover times. The other simulation (run B) has $\Omega = 9$, a Rossby 
number $\textrm{Ro} \approx 0.06$, and was continued for 30 turnover 
times. Both runs have a Reynolds number $\textrm{Re} \approx 5100$.

Figure \ref{fig:S2all} shows the result of computing the velocity 
structure functions in all directions for one snapshot of the field 
(at $t=30$ in run B), and of averaging over the different directions 
to obtain $S_2(\ell_\perp)$ and $S_2(\ell_\parallel)$. At small scales, 
both structure functions scale as $l^2$, as can be expected for a 
well-resolved 
smooth flow in the dissipative range. At intermediate scales, an inertial 
range with power law-scaling can be identified in $S_2(\ell_\perp)$, 
but not in $S_2(\ell_\parallel)$. Indeed, $S_2(\ell_\parallel)$ is 
smaller than $S_2(\ell_\perp)$ at all scales (specially so at the largest 
scales), and shows no clear scaling. This is consistent with the results 
obtained in Paper I from the energy spectrum: while the energy spectrum 
in perpendicular wave vectors shows an inertial range with power law
behavior and approximately constant flux, the energy spectrum in the 
parallel direction shows no clear scaling and its associated flux decays 
rapidly with scale.

\begin{figure}
\includegraphics[width=8.2cm]{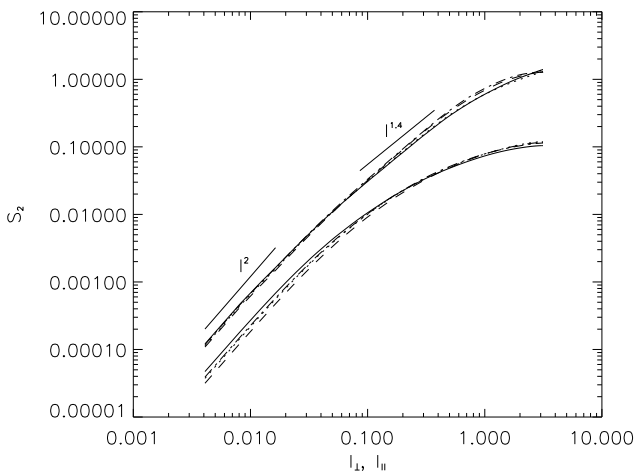}
\caption{Second order structure functions $S_2(\ell_\perp)$ and 
$S_2(\ell_\parallel)$ in run B at different times, between $t=20$ and 
$30$. A dissipative range scaling $\sim \ell^2$ is indicated at small 
scales, and the average slope $\zeta_2 \sim 1.41$ is indicated in the inertial range.}
\label{fig:S2} \end{figure}

Figure \ref{fig:S2} shows the structure functions $S_2(\ell_\perp)$ and 
$S_2(\ell_\parallel)$ at four different times between $t=20$ and $30$ in 
run B. No substantial differences are observed at the different times. The 
average scaling exponent in the inertial range 
$S_2(\ell_\perp) \sim \ell_\perp^{\zeta_2}$, obtained after averaging over 
the ten turnover times, is $\zeta_2=1.41 \pm 0.02$, and is indicated in 
the figure by the straight line. Errors here and in the following are 
defined as the standard mean error
\begin{equation}
e_{\zeta_p}= \frac{1}{N} \sqrt{\sum_{i=1}^N \left({\zeta_p}_i-
    \overline{\zeta_p}\right)^2} ,
\end{equation}
where ${\zeta_p}_i$ is the slope obtained from a least square fit 
for each snapshot $i$, $N$ is the number of snapshots, and 
$\overline{\zeta_p}$ is the mean value averaged over all snapshots. 
The error in the least square determination of the slope for each 
snapshot is much smaller than this standard mean error for the averaged 
exponent. Extended self-similarity is not used to obtain the slopes.

The value obtained for $\zeta_2$ is in good agreement with predictions 
for helical rotating turbulence \cite{Mininni09}. In a rotating flow 
with maximal helicity, $E \sim k^{-2.5}$, which leads to 
$S_2(\ell_\perp) \sim \ell_\perp^{1.5}$. For flows with non-maximal 
helicity the $\zeta_2$ exponent is, according to \cite{Mininni09}, 
between 1 and $1.5$, with the value of $1$ corresponding to the 
non-helical case. Note that in numerical simulations of non-helical 
rotating turbulence $S_2(\ell_\perp) \sim \ell_\perp$ was reported in 
\cite{Muller07,Mininni09b}.

\begin{figure}
\includegraphics[width=8.2cm]{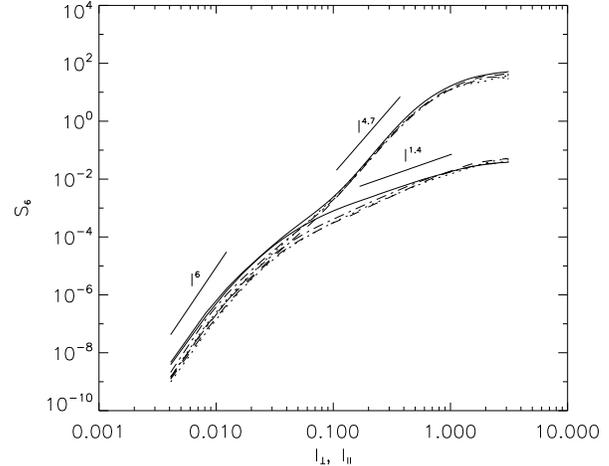}
\caption{Sixth order structure functions $S_6(\ell_\perp)$ and 
$S_6(\ell_\parallel)$ in run B at different times, between $t=20$ and 
$30$. A dissipative range scaling $\sim \ell^6$ is indicated at small 
scales, and two average slopes are indicated in the inertial ranges (see text). Note that the perpendicular part of the structure function dominates the parallel one at all scales.}
\label{fig:S6} \end{figure}

From the behavior of the amplitude of $S_2(\ell_\parallel)$ with  
scale, it seems isotropy could be recovered at small scales in a 
rotating flow if the inertial range is wide enough for $S_2(\ell_\parallel)$ 
and $S_2(\ell_\perp)$ to collapse (or, in other words, for the effect 
of rotation to be negligible at the smallest scales). Indeed, for small 
values of $\ell$ the parallel and perpendicular structure functions 
become closer. This is more evident in velocity structure functions of 
higher order. As an example, Figure \ref{fig:S6} shows the sixth order 
velocity structure function. While an inertial range in the 
perpendicular direction is still visible (the mean slope averaged over 
ten turnover times is indicated as a reference), and the structure 
functions in this direction do not change much between different 
snapshots, such is not the case at smaller scales, or in the parallel 
direction. In the parallel direction, larger fluctuations between different times 
are observed. Moreover, the anisotropic inertial range is shortened as, 
at small scales, both $S_6(\ell_\perp)$ and $S_6(\ell_\parallel)$ 
collapse to a single curve. This collapse takes place for scales 
smaller than $\ell \approx 0.05$, with slight fluctuations in 
time; it indicates that the strongest events in the flow 
(which begin to dominate structure functions as the order increases) 
 tend towards 
isotropy at the smallest scales. Note that the collapse is not observed 
in the $S_2$ structure functions (see Fig. \ref{fig:S2}), whereas 
$S_p(\ell_\perp)$ and $S_p(\ell_\parallel)$ get closer to each other at smaller scales 
as the order $p$ is increased.

\begin{figure}
\includegraphics[width=8.2cm]{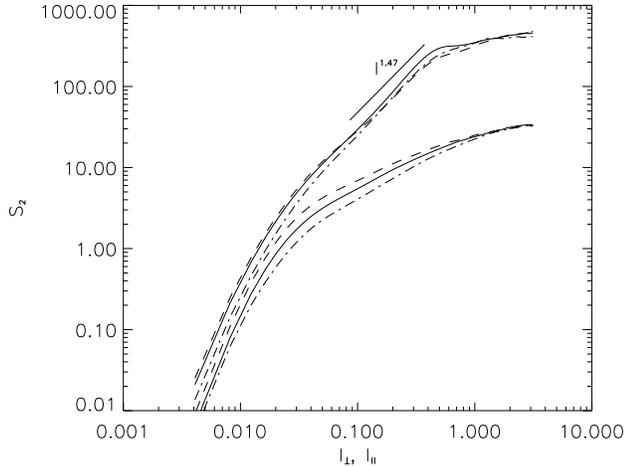}
\caption{Second-order helicity structure functions $H_2(\ell_\perp)$ and 
$H_2(\ell_\parallel)$ (see Eq. (\ref{eq:structureh}))
 in run B with $Ro=0.06$. at different times between $t=20$ and 
$30$. The dissipative range scales as $\sim \ell^4$, 
consistent with the fact that $H_2$ is quartic in the velocity; 
the average slope 
is indicated for the inertial range.}
\label{fig:H2} \end{figure}

This partial recovery of isotropy in the higher order moments of the 
velocity field can be understood as follows: higher values of $p$ in 
Eq. (\ref{eq:structurev}) leave only the strongest gradients 
contributing to $S_p(\ell_\perp)$ and $S_p(\ell_\parallel)$, as the 
contribution of mild gradients to the structure functions goes to zero 
as $p$ is increased. It is only these strong events at small scales 
that are more isotropically distributed, as indicated by Fig. 
\ref{fig:S6} (at least at the Reynolds numbers considered here). However, 
mild gradients are still anisotropic at small scales, as follows from Fig. 
\ref{fig:S2}. Simulations with more spatial resolution at larger Reynolds 
numbers will be required to study if this transition towards isotropy only 
takes place for the strong events, or whether for sufficiently small 
scales the transition takes place for all orders, with a change in the 
spectral index of the flow (see e.g., \cite{Dubrulle92,Zeman94,Zhou95} 
for phenomenological treatments of non-helical rotating turbulence that 
consider the possibility of such a transition). 

Similarly to the stratified case, one can introduce a wavenumber at which rotation and nonlinear advection balance (see \cite{Dubrulle92,Zeman94}):
$$k_\Omega = (\Omega^3/\epsilon)^{1/2}$$
 (with $\epsilon$ the energy dissipation rate);
$k_{\Omega}$ can be considered as the largest wavenumber 
where rotation effects are important; the inverse of this wavenumber, 
$\ell_\Omega = 2\pi/k_\Omega $, 
is equivalent to the Ozmidov length in 
stratified turbulence, which separates the inertial range (at smaller 
scales) from the range dominated by buoyancy (at larger scales). It is 
rather remarkable that in our simulation $\ell_\Omega \approx 0.04$, 
close to the value of $\ell \approx 0.05$ where the transition in 
$S_p(\ell_\perp)$ takes place for large values of $p$ (see Fig. 
\ref{fig:S6}). However, a confirmation of this would require a 
parametric study varying the value of $\Omega$, which in DNS at the 
spatial resolution considered here is out of reach with present day 
computers.

At this point, a discussion about units is in order. The distance between 
grid points in our runs is $2\pi/1536 \approx 0.004$. This is also the 
smallest distance for which increments can be computed in the structure 
functions. Since the simulations are dealiased using the 2/3-rule, the 
largest wavenumber resolved is $k_\textrm{max} = 512$ which corresponds 
to a length $\ell_\textrm{min} = 2\pi/k_\textrm{max} \approx 0.01$. As 
a result, the velocity field at scales between $\approx 0.004$ and 
$\approx 0.01$ must be necessarily smooth, and its structure functions 
should scale as $S_p \sim \ell^p$ as it is indeed the case (see e.g., 
Fig. \ref{fig:S2}). The dissipation scale in the simulations is just 
slightly larger than $\ell_\textrm{min}$, which explains why the 
$\sim \ell^p$ scaling extends a little bit beyond $\ell_\textrm{min}$ (as 
required for the simulations to be well resolved). For practical 
purposes, we can estimate the dissipation scale to be between $\approx 0.01$ 
and $0.02$ (these values are consistent with estimations from the energy 
spectrum, shown in Paper I). Since the collapse of the parallel and 
perpendicular structure functions occurs near $\ell \approx 0.05$, one may 
wonder if this scale is well resolved, or if the collapse results from 
numerical cut-off or viscous effects. We computed structure functions for 
simulations of forced helical and non-helical rotating turbulence at 
resolutions of $512^3$ grid points (see \cite{Mininni09b,Mininni09}) for 
which $k_\Omega > k_\textrm{max}$, and in that case no bump, collapse, or 
changes in the behavior of the structure functions at the smallest scales were 
observed. However, simulations at larger resolutions would be desirable 
to further confirm this result.

\subsection{Helicity structure functions} 

\begin{figure}
\includegraphics[width=8.2cm]{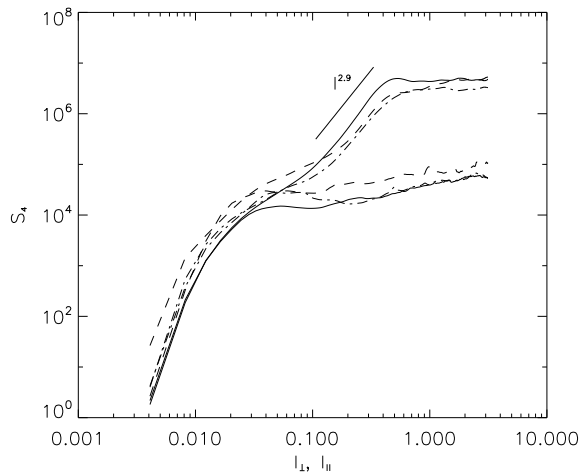}
\caption{Fourth-order helicity structure functions $H_4(\ell_\perp)$ and 
$H_4(\ell_\parallel)$ in run B at different times, between $t=20$ and 
$30$. The average slope is indicated for the inertial range.}
\label{fig:H4} \end{figure}

Helicity structure functions are computed in the same fashion as velocity 
structure functions (see Eq. (\ref{eq:structureh})); 
the functions $H_2(\ell_\perp)$ and $H_2(\ell_\parallel)$ 
for different times are shown in Fig. \ref{fig:H2}, after averaging in the 
different directions. It should be noted that the $H_2$ structure 
functions are effectively of fourth order in the fields, and as a result 
the convergence of the statistics is not as good in this case as it was 
for the structure functions discussed in the previous section for the velocity. Also as a 
result of the higher-order dependence on the fields, the dissipative range 
scales as $\ell^4$. Moreover, helicity is not a positive definite 
quantity, and cancellations between regions with positive and negative 
alignments of the velocity and the vorticity can take place, resulting 
in larger fluctuations of the increments. Convergence of the statistics 
for all orders studied here was checked by computing the cumulants for 
each moment (see e.g., \cite{Gotoh02,Mininni08}).

\begin{figure}
\includegraphics[width=8.2cm]{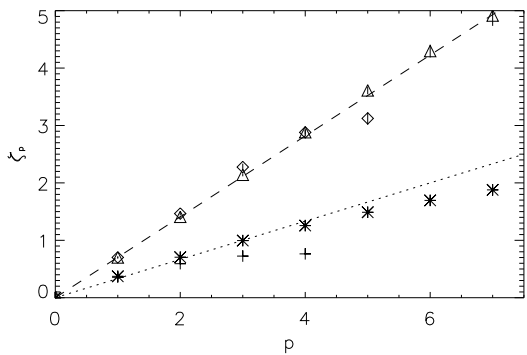}
\caption{Scaling exponents (with error bars, see Table \ref{table:exponents})
as a function of the order 
$p$, for the velocity (stars) and the helicity (pluses) in run A with $Ro=8.5$, and 
for the velocity (triangles) and the helicity (diamonds) in run B with $Ro=0.06$. The 
dotted line corresponds to Kolmogorov scaling $\zeta_p=p/3$, and the dash
line to $\zeta_p=0.71 p$, which represents the velocity exponents best.}
\label{fig:exponents} \end{figure}

In the case of the helicity structure functions, fluctuations between 
different temporal snapshots are larger than for the velocity structure 
functions. This is mostly due to the fact that $H_2(\ell_\parallel)$ 
displays strong fluctuations in time, and that the helicity becomes 
isotropic at smaller scales more rapidly with increasing $p$. However, 
fluctuations in the inertial range of the perpendicular increments are 
smaller, and an inertial range can still be defined. For the second 
order, defining the scaling exponents as 
${\cal H}_p(\ell_\perp) \sim \ell_\perp^{\xi_p}$, we obtain 
$\xi_2 = 1.46 \pm 0.03$. This value leads to a spectral scaling 
$H(k) \sim k_\perp^{-1.7}$ (the prediction for a maximally-helical rotating 
flow is $H(k) \sim k_\perp^{-1.5}$). The scaling of the energy and of the 
helicity obtained from the structure functions is consistent with the 
$e+h=4$ rule for helical rotating flows, where $e$ is the energy 
spectral index and $h$ is the helicity spectral index \cite{Mininni09};
it is also consistent with the spectral indices measured in Paper I.

Fluctuations become larger for larger values of $p$, and we were unable 
to compute structure functions for moments with converged cumulants 
beyond $p=4$ for run A and $p=5$ for run B for the helicity. Specifically for run B, 
variations in the amplitude of ${\cal H}_p(\ell_\parallel)$ increase with $p$, 
changing the scale where the structure functions become isotropic. As a 
result, for some snapshots no scaling in the perpendicular direction 
was observed, and as a rule of thumb snapshots for which 
${\cal H}_p(\ell_\perp)$ and ${\cal H}_p(\ell_\parallel)$ became of the same order 
at scales larger than $\ell \approx 0.1$ had to be discarded. Figure 
\ref{fig:H4} shows the fourth order helicity structure functions for 
four snapshots that present a discernible inertial range in 
$\ell_\perp$ (a total of eight snapshots was used to compute the 
average scaling exponents). 

\begin{table}
\caption{\label{table:exponents} Order $p$, and scaling exponents 
$\zeta_p$ for the velocity and $\xi_p$ for the helicity, with errors, for run A ($Ro=8.5$)  and run B ($Ro=0.06$).}
\begin{ruledtabular}
\begin{tabular}{ccccc}
p  &$\zeta_p$ (run A)&$\xi_p$ (Run A) & 
    $\zeta_p$ (run B)&$\xi_p$ (Run B) \\
\hline
1  &$0.37\pm0.01$&$0.35\pm0.01$&$0.701\pm0.007$&$0.70\pm0.03$ \\
2  &$0.70\pm0.03$&$0.59\pm0.01$&$1.41\pm0.02$  &$1.46\pm0.03$ \\
3  &$0.99\pm0.03$&$0.72\pm0.01$&$2.14\pm0.02$  &$2.27\pm0.07$ \\
4  &$1.25\pm0.04$&$0.76\pm0.02$&$2.88\pm0.04$  &$2.88\pm0.07$ \\
5  &$1.49\pm0.04$& ---         &$3.61\pm0.07$  &$3.1\pm0.1$   \\
6  &$1.69\pm0.05$& ---         &$4.3\pm0.1$    & ---          \\
7  &$1.88\pm0.05$& ---         &$4.9\pm0.2$    & ---          \\
\end{tabular}
\end{ruledtabular}
\end{table}

\begin{figure*}
\includegraphics[width=16cm]{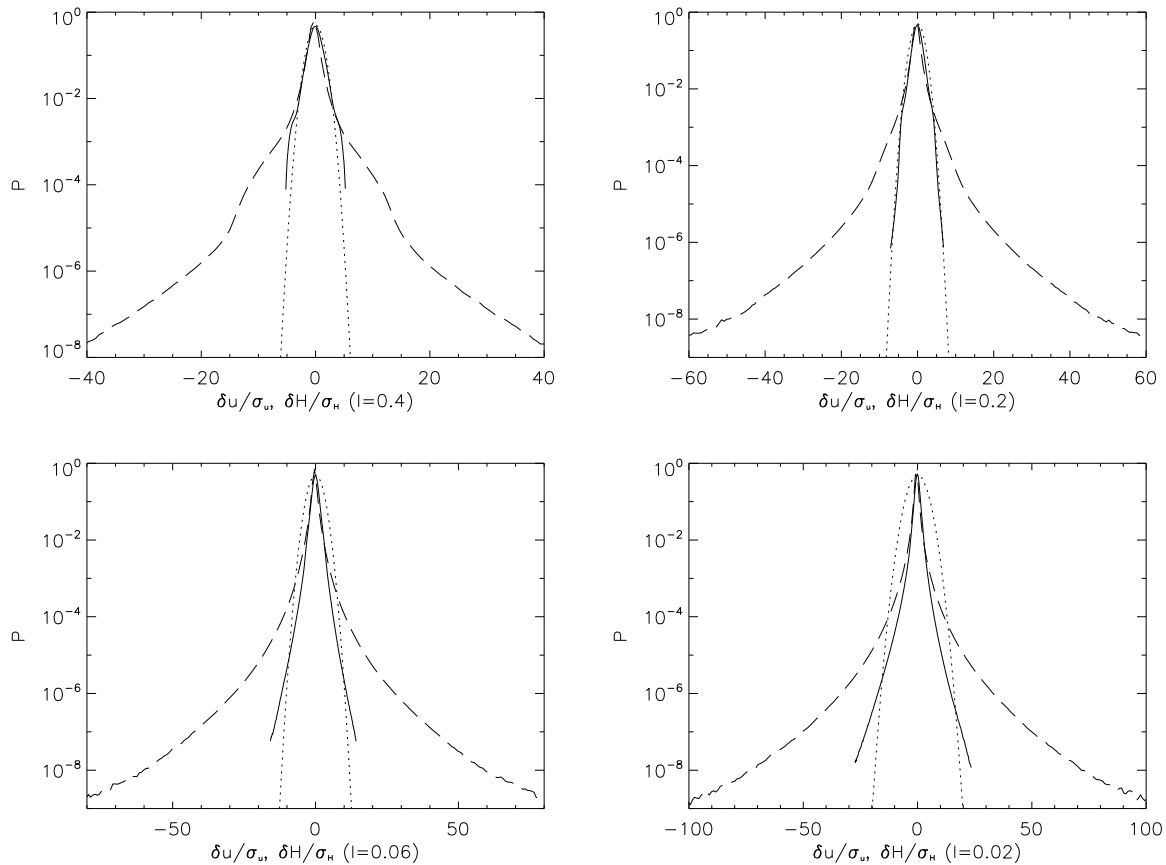}
\caption{Probability density functions at different intervals in the direct cascade 
for velocity (solid) and helicity 
(dashed) increments in the direction perpendicular to the axis of 
rotation. Increments are normalized by their variance. The dotted curve 
represents a Gaussian distribution with the same variance.}
\label{fig:histo} \end{figure*}

\section{Intermittency in the direct cascades\label{sec:intermittency}}

In a self-similar flow, scaling exponents depend linearly on the order 
$p$. As mentioned in the introduction, the anomalous scaling (the 
deviation from linear dependence on $p$) of the exponents observed in 
many turbulent flows is a signature of deviations from scale invariance 
and of intermittency. The velocity and helicity scaling exponents in 
the direct cascade range of runs A and B are shown in Fig. 
\ref{fig:exponents}.

In run A, with $\Omega=0.06$, the effect of rotation is negligible 
and the velocity scaling exponents display the usual deviation from the 
Kolmogorov $p/3$ scaling. Deviations from a the straight line are often 
quantified in terms of the intermittency coefficient 
$\mu = 2\zeta_3 - \zeta_6$, which for this run is $\mu = 0.29 \pm 0.06$, 
in agreement with previous simulations and experiments of non-rotating 
turbulence. The third order exponent is $\zeta_3=0.99\pm0.03$, in good 
agreement with the value of 1 expected for isotropic and homogeneous 
turbulence. The higher orders computed in this run are also consistent 
with results of non-rotating turbulence at very large Reynolds numbers 
(see e.g., \cite{Mininni08}). All values of the scaling exponents up to 
order 8 are given in Table \ref{table:exponents}. 

On the other hand, for run B at low Rossby number, the velocity scaling exponents are (within 
error bars) consistent with a scale invariant (intermittency-free) linear relationship. For this run, $\mu = 0.0 \pm 0.1$, compatible with a 
value of $\mu=0$ which corresponds to a completely scale invariant flow. 
A similar result was reported in an experimental study \cite{Baroud03} 
(although associated in that case with a linear scaling $\zeta_p=p/2$ and in a range of scales that may correspond to an inverse cascade of energy). 
The values of $\zeta_p$ for run B contrast with results obtained 
for the scaling exponents in non-helical rotating turbulence from DNS for the direct cascade 
\cite{Muller07,Mininni08} and from laboratory experiments for decaying flows \cite{Seiwert08}, 
where a reduction of the intermittency was observed but anomalous scaling 
as a signature of intermittency was still present; as an example, in Ref. 
\cite{Mininni08}, for a non-helical flow at late times with 
${\textrm Ro}\approx 0.07$ it was found that $\mu = 0.24\pm0.02$. The 
possibility that the different behaviors reported in the experiments 
can be ascribed to helicity is tantalizing.

In the case of the helicity exponents $\xi_p$, the highest orders could 
not be measured for the reasons discussed above: the higher order 
dependence on the fields of the helical structure functions, and the 
associated stronger fluctuations observed. However, results for run A 
are in agreement with previous studies of the direct cascade of helicity 
in isotropic and homogeneous turbulence, with the helicity being more 
intermittent than the velocity field (note however differences in the 
definitions used here and in the analysis in \cite{Chen03b}). This is 
illustrated by the smaller values of $\xi_p$ when compared with $\zeta_p$ 
in run A (see Fig. \ref{fig:exponents} and Table \ref{table:exponents}). 
On the other hand, results in run B at low Rossby number seem to be of a different nature. 
The $\xi_p$ exponents up to $p=4$ are within error bars consistent with 
a linear (self-similar) scaling $\xi_p \approx 0.73 p$ (i.e., with 
a slightly larger slope than for $\zeta_p$ in the same run, see Fig. 
\ref{fig:exponents}), but $\xi_5$ departs from such scaling (see also 
Table \ref{table:exponents}). This departure would indicate 
intermittency in the helicity, a property that will be confirmed in the 
next section studying the probability density functions of the increments.

\begin{figure}
\includegraphics[width=8.5cm]{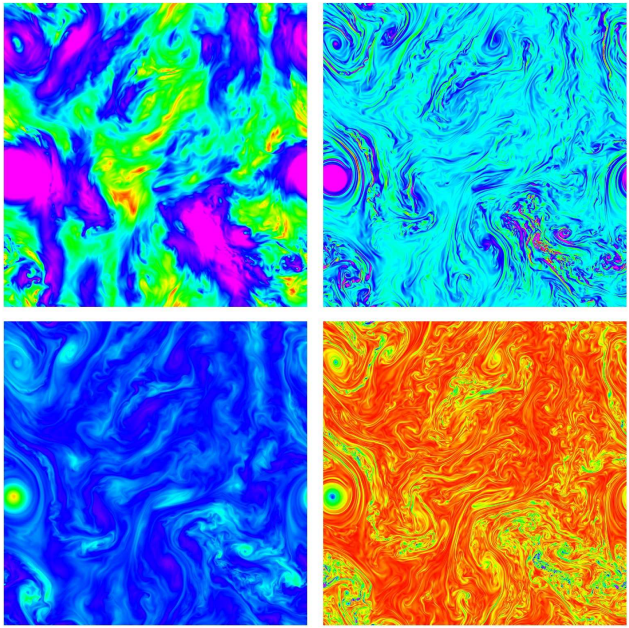}
\caption{(Color online) Slices of the energy density (top left), vorticity intensity (top right), $z$ component of the velocity (bottom left), and helicity density (bottom right), in run B at $t \approx 30$. Note the imprint of small scales in the vorticity and helicity (right column).}
\label{fig:2D} \end{figure}

\begin{figure}
\includegraphics[width=8cm]{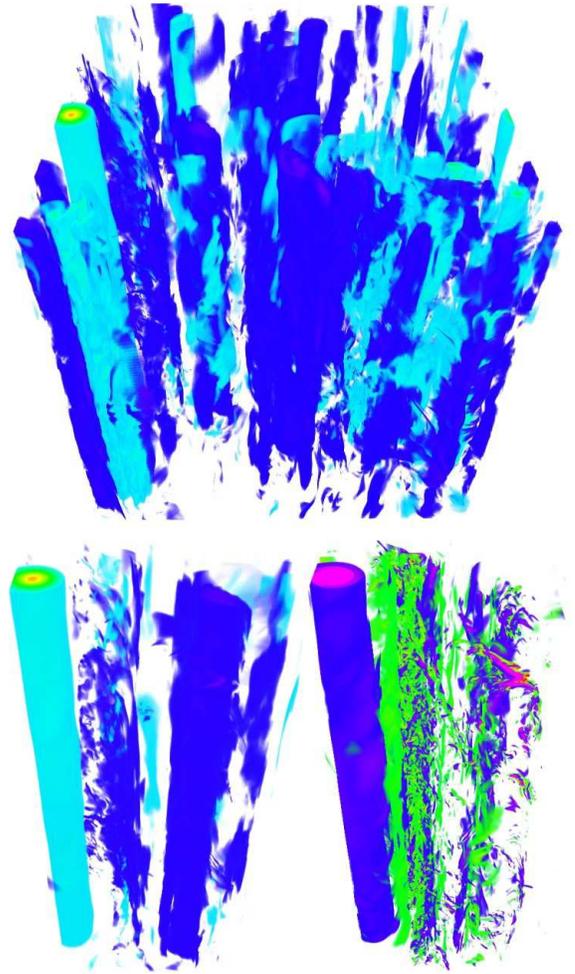}
\caption{(Color online) Three dimensional rendering of the $z$ component of the velocity in the entire domain in run B at $t \approx 30$ (above), and a zoom on a subregion (below) showing the $z$ component of the velocity in a column-like structure (left) and its helicity density (right).}
\label{fig:3D} \end{figure}

\section{Probability density functions \label{sec:histo}}

The identification of multi-fractal (as opposed to scale independent) 
scaling in turbulent flows based on scaling exponents present several 
difficulties. It is well known that transients and finite size effects 
can give spurious multi-fractal scaling \cite{Aurell97}, and that 
logarithmic or sub-leading corrections to the spectrum can also lead to 
the same result \cite{Mitra05}. In this section we consider probability 
density functions of the velocity and helicity increments. In a scale 
invariant flow, the velocity increments are expected to be Gaussian. 
On the other hand, in an intermittent flow, probability density functions 
are expected to have strong non-Gaussian tails.

In Fig. \ref{fig:histo} we show the probability density functions (PDFs) 
of the velocity and helicity increments in the direction perpendicular 
to the axis of rotation for four values of $\ell_\perp = 0.4$, $0.2$, 
$0.06$, and $0.02$. The PDFs are normalized by their variance, and a 
Gaussian with the same variance is shown as a reference. The PDFs of 
velocity increments for $\ell_\perp = 0.4$ and $0.2$ are close to 
Gaussian. Note that these increments correspond to scales in the 
anisotropic inertial range (see Figs. \ref{fig:S2} and \ref{fig:S6}). 
PDFs close to Gaussian were also observed for $\delta u$ for other 
increments in this range. However, for $\ell_\perp = 0.4$ a ``bump'' 
(which also decays as a Gaussian) can be observed in the tails. This 
bump disappears as smaller increments are considered, and seems to be 
associated with the forcing (note that the mechanical forcing acts 
between $\ell \approx 0.78$ and $0.90$).

The bump is more evident in the PDF of helicity increments with 
$\ell_\perp = 0.4$. For both values $\ell_\perp = 0.4$ and $0.2$ (in 
the direct cascade inertial range) the PDFs of helicity increments are 
different than for velocity increments: deviations from Gaussianity are 
evident, and the PDFs show strong tails. The non-Gaussian tails are the 
signature of the presence of strong gradients, and of intermittency in 
the spatial distribution of helicity. These results confirm -- independently 
of the scaling exponents analyzed in the preceding section -- 
that the anisotropic direct 
cascade of energy towards smaller scales is close to Gaussian and scale 
independent, while the direct cascade of helicity is intermittent.

For increments in the range of scales where $S_p(\ell_\perp)$ is of 
the same order as $S_p(\ell_\parallel)$ for $p \ge 4$ (see e.g., Fig. 
\ref{fig:S6}), both the PDFs of velocity and of helicity increments 
show non-Gaussian tails (see Fig. \ref{fig:histo} for $\ell_\perp=0.06$ 
and $0.02$). This further confirms that at very small scales, the 
strongest events (in the tails of the PDFs) tend toward a recovery of 
isotropy, which would lead in turn to a traditional (and intermittent) 
direct cascade of energy. However, simulations at larger resolution and 
at different Reynolds and Rossby numbers will be required to verify if 
this recovery of isotropy occurs always near the dissipation range, or 
if a second isotropic inertial range develops when enough scale 
separation is available at scales smaller than the Ozmidov scale when rotation can presumably be neglected.

\section{Structures\label{sec:structures}}

The results discussed in Sect. \ref{sec:intermittency} and \ref{sec:histo} 
point towards an anisotropic and scale-invariant energy distribution in 
an intermediate range of scales (smaller than the forcing scale), and 
a highly intermittent helicity distribution in the same range of scales. 
In helical rotating turbulence, energy cascades both towards large 
and small scales, while helicity cascades towards small scales 
dominating the direct cascade inertial range. In light of these 
facts, it is of interest to look at the structures that 
arise in the flow.

In Figure \ref{fig:2D} are displayed horizontal slices (in the plane perpendicular to the 
rotation axis) of the energy density, the vorticity intensity, the 
$z$ component of the velocity, and the helicity density in run B with $Ro=0.06$, at 
$t\approx 30$. While the energy density seems organized in large-scale 
patches (substantially larger than the forcing scale which corresponds 
roughly to $1/7$ of the box), the vorticity intensity and helicity 
density show small-scale structures. This is consistent with an inverse 
cascade of energy and a direct cascade dominated by the helicity. 
Note however that the energy density distribution in space is different 
from the distribution observed in two dimensional turbulence, where an 
inverse cascade of energy also takes place;
in other words, the inverse cascade here differs from the purely two-dimensional case, be it only because the conservation of helicity (in the ideal case) induces the flow to keep some trace of three-dimensionality (and isotropy) at small scale, as noted before. 
Moreover, smooth structures at some intermediate 
scale can be observed in the helicity and the 
vorticity (see e.g. the left side of the box); these regions are also 
correlated with similar regions in the $z$ component of the velocity.

When run B is started from a previously isotropic state, the flow 
first becomes anisotropic and then a self-organization process starts 
that leads to the formation of columns. Those columns can be identified 
when the energy density, helicity density, or the vorticity intensity 
are visualized. The columns have strong $u_z$ (see e.g. Fig. 
\ref{fig:3D}). However, a few columns can be distinguished from the rest, 
in that they have a strong updraft velocity and concentrate in their 
core positive helicity, with strong relative helicity (strong alignment
between velocity and vorticity).
 These columns are stable, and we were able to 
track these columns in the simulation for over ten turnover times.

Far from these structures, the flow displays a myriad of small scales, 
as illustrated by the small-scale filaments in the helicity density. 
These filaments are also organized in columnar structures, but the 
thick columns with net helicity live for much longer times. As a result, 
in real space the inverse cascade of energy can be identified as the 
system evolves in time and these helical structures merge with 
columns with a lesser amount of relative helicity, 
increasing the characteristic width of the column when its 
energy density is visualized, but keeping the thickness of the helical 
core in the column approximately constant. This is the result of the 
helicity injected in the flow cascading directly to smaller scales, 
which allows for a localized helical column, but prevents the formation 
of a thick distribution of helicity in a column filling all space. 
Accompanying the direct cascade, strong fluctuations of helicity are 
observed in the turbulent columns with the characteristic size of the 
vortex filaments.

\section{Conclusions\label{sec:conclusions}}

The analysis of the structure functions and of the probability density 
functions of velocity and helicity increments from data stemming from 
direct numerical simulations of helical rotating turbulence at high resolution, showed that, at least for the strongest events in the small 
scales, isotropy is recovered at sufficiently small scales. This is 
observed both in the collapse of the parallel and perpendicular structure 
functions, as well as in the transition from near-Gaussian statistics of 
the velocity increments in the inertial range towards PDFs with strong 
tails in the same range of scales as observed in the structure functions. 
More studies will be required to see if this transition takes place for 
all orders when the Reynolds number is large enough, leading to  a sufficient scale separation.

Concerning scaling exponents, we also confirmed that in the anisotropic 
direct cascade range the velocity increments are (within error bars) 
scale invariant (i.e., non-intermittent) while helicity increments are 
intermittent. This is further confirmed by the probability density 
functions, which show strong non-Gaussian tails for the helicity, and 
are near Gaussian for velocity increments in the anisotropic range. The 
scaling exponents for the helicity are consistent (within error bars) 
with a scale invariant dependence $\xi_p \approx 0.73 p$ up to $p=4$, but 
$\xi_5$ departs from a straight line. More data will be required to 
confirm a possible bi-fractal or multi-fractal scaling for the helicity, 
although we would like to point out that a bi-fractal scaling would be 
consistent with the two types of helical structures observed in 
visualizations of the flow (the large-scale laminar columns and the 
small-scale vortex filaments), and reminiscent of the behavior or other 
systems with two type of structures, e.g., the Burgers equation, which 
develops smooth ramps connected by sharp shocks.

\begin{acknowledgments}
Computer time was provided by NCAR. NCAR is sponsored by the National 
Science Foundation. PDM acknowledges support from grant UBACYT X468/08 
and PICT-2007-02211, and from the Carrera del Investigador Cient\'{\i}fico 
of CONICET.
\end{acknowledgments}


\begin{thebibliography}{10}

\bibitem{Frisch}
U.~Frisch, {\em Turbulence: the legacy of A.N. Kolmogorov} 
(Cambridge Univ.\ Press, Cambridge, 1995).

\bibitem{Sreenivasan97}
K.~R. Sreenivasan and R.~A. Antonia, ``The phenomenology of small-scale
  turbulence,'' Annu.\ Rev.\ Fluid Mech. {\bf 29}, 437 (1997).

\bibitem{Sorriso00}
L.~Sorriso-Valvo, V.~Carbone, P.~Veltri, H.~Politano, and A.~Pouquet,
  ``Non-gaussian probability distribution functions in two-dimensional
  magnetohydrodynamic turbulence,'' Europhys.\ Lett. {\bf 51}, 520 (2000).

\bibitem{Mininni07}
P.~D. Mininni and A.~Pouquet, ``Energy spectra stemming from interactions of
  {A}lfv\'en waves and turbulent eddies,'' Phys.\ Rev.\ Lett. {\bf 99},
  254502 (2007).

\bibitem{Bernard06}
D.~Bernard, G.~Boffetta, A.~Celani, and G.~Falkovich, ``Conformal invariance in
  two-dimensional turbulence,'' Nature Physics {\bf 2}, 124 (2006).

\bibitem{Cambon89}
C.~Cambon and L.~Jacquin, ``Spectral approach to non-isotropic turbulence
  subjected to rotation,'' J.\ Fluid Mech. {\bf 202}, 295 (1989).

\bibitem{Waleffe93}
F.~Waleffe, ``Inertial transfers in the helical decomposition,'' Phys.\
  Fluids A {\bf 5}, 677 (1993).

\bibitem{Smith05}
L.~M. Smith and Y.~Lee, ``On near resonances and symmetry breaking in forced
  rotating flows at moderate rossby number,'' J.\ Fluid Mech. {\bf 535},
  111 (2005).

\bibitem{Seiwert08}
J.~Seiwert, C.~Morize, and F.~Moisy, ``On the decrease of intermittency in
  decaying rotating turbulence,'' Phys.\ Fluids {\bf 20}, 071702 (2008).

\bibitem{Muller07}
W.-C. M\"uller and M.~Thiele, ``Scaling and energy transfer in rotating
  turbulence,'' Europhys.\ Lett. {\bf 77} 34003 (2007).

\bibitem{Mininni09b}
P.~D. Mininni, A.~Alexakis, and A.~Pouquet, ``Scale interactions and scaling
  laws in rotating flows at moderate {R}ossby numbers and large {R}eynolds
  numbers,'' Phys.\ Fluids {\bf 21}, 015108 (2009).

\bibitem{Baroud03}
C.~N. Baroud, B.~B. Plapp, H.~L. Swinney, and Z.-S. She, ``Scaling in
  three-dimensional and quasi-two-dimensional rotating turbulent flows,'' 
  Phys.\ Fluids {\bf 15}, 2091 (2003).

\bibitem{Hoyng93}
P.~Hoyng, ``Helicity fluctuations in mean field theory: an explanation for the
  variability of the solar cycle?,'' Astron.\ Astrophys. {\bf 272},
  321 (1993).

\bibitem{Charbonneau01}
P.~Charbonneau, ``Multiperiodicity, chaos, and intermittency in a reduced model
  of the solar cycle,'' Sol.\ Phys. {\bf 199}, 385 (2001).

\bibitem{Mininni04}
P.~D. Mininni, D.~O., and G\'omez, ``A new technique for comparing solar dynamo
  models and observations,'' Astron.\ Astrophys. {\bf 426}, 1065 (2004).

\bibitem{Kulkarni99}
J.~R. Kulkarni, L.~K. Sadani, and B.~S. Murthy, ``Wavelet analysis of
  intermittent turbulent transport in the atmospheric surface layer over a
  monsoon trough region,'' Bound.-Layer Meteor. {\bf 90}, 217 (1999).

\bibitem{paper1}
P.~Mininni and A.~Pouquet, ``Rotating helical turbulence. {P}art {I}. {G}lobal
  evolution and spectral behavior.'' submitted to Phys. of Fluids (2009).

\bibitem{Lilly88}
D.~K. Lilly, ``The structure, energetics, and propagation of rotating
  convective storms. {P}art {I}{I}: helicity and storm stabilization,''
  J.\ Atmosph.\ Sc. {\bf 43}, 126 (1988).

\bibitem{Kerr96}
B.~W. Kerr and G.~L. Darkow, ``Storm-relative winds and helicity in the
  tornadic thunderstorm environment,'' Weath.\ and Forecast. {\bf 11},
  489 (1996).

\bibitem{Markowski98}
P.~M. Markowski, J.~M. Straka, E.~N. Rasmussen, and D.~O. Blanchard,
  ``Variability of storm-relative helicity during {V}{O}{R}{T}{E}{X},''
  Mont.\ Weath.\ Rev. {\bf 126}, 2959 (1998).

\bibitem{Mininni09}
P.~D. Mininni and A.~Pouquet, ``Helicity cascades in rotating turbulence,''
  Phys.\ Rev.\ E {\bf 79}, 026304 (2009).

\bibitem{Kolmogorov41}
A.~N. Kolmogorov, ``Dissipation of energy in locally isotropic turbulence,''
  Dokl.\ Akad.\ Nauk SSSR {\bf 32}, 16 (1941).

\bibitem{Chen03}
Q.~Chen, S.~Chen, and G.~L. Eyink, ``The joint cascade of energy and helicity
  in three-dimensional turbulence,'' Phys.\ Fluids {\bf 15}, 361 (2003).

\bibitem{Chen03b}
Q.~Chen, S.~Chen, G.~L. Eyink, and D.~D. Holm, ``Intermittency in the joint
  cascade of energy and helicity,'' Phys.\ Rev.\ Lett. {\bf 90},
  214503 (2003).

\bibitem{Chkhetiani96}
O.~G. Chkhetiani, ``On the third-moments in helical turbulence,'' JETP
  Lett. {\bf 63}, 768 (1996).

\bibitem{Kurien03}
S.~Kurien, ``The reflection-antisymmetric counterpart of the
  {K}\'arm\'an-{H}owarth dynamical equation,'' Phys.\ D {\bf 175},
  167 (2003).

\bibitem{Gomez00}
T.~Gomez, H.~Politano, and A.~Pouquet, ``Exact relationship for third-order
  structure functions in helical flows,'' Phys.\ Rev.\ E {\bf 61},
  5321 (2000).

\bibitem{Tsinober92}
A.~Tsinober, E.~Kit, and T.~Dracos, ``Experimental investigation of the field
  of velocity gradients in turbulent flows,'' J.\ Fluid Mech. {\bf 242},
  169 (1992).

\bibitem{Jacquin90}
L.~Jacquin, O.~Leuchter, C.~Cambon, and J.~Mathieu, ``Homogeneous turbulence in
  the presence of rotation,'' J.\ Fluid Mech. {\bf 220}, 1 (1990).

\bibitem{Morize05}
C.~Morize, F.~Moisy, and M.~Rabaud, ``Decaying grid-generated turbulence in a
  rotating tank,'' Phys.\ Fluids {\bf 17}, 095105 (2005).

\bibitem{Staplehurst08}
P.~J. Staplehurst, P.~A. Davidson, and S.~B. Dalziel, ``Structure formation in
  homogeneous freely decaying rotating turbulence,'' J.\ Fluid Mech.,
  {\bf 598}, 81 (2008).

\bibitem{Smith99}
L.~M. Smith and F.~Waleffe, ``Transfer of energy to two-dimensional large
  scales in forced, rotating three-dimensional turbulence,'' Phys.\
  Fluids {\bf 11}, 1608 (1999).

\bibitem{Cambon97}
C.~Cambon, N.~N. Mansour, and F.~S. Godeferd, ``Energy transfer in rotating
  turbulence,'' J.\ Fluid Mech. {\bf 337}, 303 (1997).

\bibitem{Davidson06}
P.~A. Davidson, P.~J. Staplehurst, and S.~B. Dalziel, ``On the evolution of
  eddies in a rapidly rotating system,'' J.\ Fluid Mech. {\bf 557},
  135 (2006).

\bibitem{Arad98}
I.~Arad, B.~Dhruva, S.~Kurien, V.~S. L'vov, I.~Procaccia, , and K.~R.
  Sreenivasan, ``Extraction of anisotropic contributions in turbulent flows,''
  Phys.\ Rev.\ Lett. {\bf 81}, 5330 (1998).

\bibitem{Biferale05}
L.~Biferale and I.~Procaccia, ``Anisotropy in turbulent flows and in turbulent
  transport,'' Phys.\ Rep. {\bf 414}, 43 (2005).

\bibitem{Taylor03}
M.~A. Taylor, S.~Kurien, and G.~L. Eyink, ``Recovering isotropic statistics in
  turbulence simulations: the {K}olmogorov 4/5th law,'' Phys.\ Rev.\ E,
  {\bf 68}, 026310 (2003).

\bibitem{Dubrulle92}
B.~Dubrulle and L.~Valdettaro, ``Consequences of rotation in energetics of
  accretion disks,'' Astron.\ Astrophys. {\bf 263}, 387 (1992).

\bibitem{Zeman94}
O.~Zeman, ``A note on the spectra and decay of rotating homogeneous
  turbulence,'' Phys.\ Fluids {\bf 6}, 3221 (1994).

\bibitem{Zhou95}
Y.~Zhou, ``A phenomenological treatment of rotating turbulence,'' Phys.\
  Fluids {\bf 7}, 2092 (1995).

\bibitem{Gotoh02}
T.~Gotoh, D.~Fukayama, and T.~Nakano, ``Velocity field statistics in
  homogeneous steady turbulence obtained using a high-resolution direct
  numerical simulation,'' Phys.\ Fluids {\bf 14}, 1065 (2002).

\bibitem{Mininni08}
P.~D. Mininni, A.~Alexakis, and A.~Pouquet, ``Nonlocal interactions in
  hydrodynamic turbulence at high {R}eynolds numbers: The slow emergence of
  scaling laws,'' Phys.\ Rev.\ E {\bf 77}, 036306 (2008).

\bibitem{Aurell97}
E.~Aurell, U.~Frisch, A.~Noullez, and M.~Blank, ``Bifractality of the devil's
  staircase appearing in the {B}urgers equation with brownian initial
  velocity,'' J.\ Stat.\ Phys. {\bf  88}, 1151 (1997).

\bibitem{Mitra05}
D.~Mitra, J.~Bec, R.~Pandit, and U.~Frisch, ``Is multiscaling an artifact in
  the stochastically forced {B}urgers equation?,'' Phys.\ Rev.\ Lett.,
  {\bf 94}, 194501 (2005).

\end{thebibliography}

\end{document}